# Sensitivity of Isothermal Swirl Combustor Flow to Inlet Reynolds Number

Madan Lal Mahato[1], Nitesh Kumar Sahu[1*]

[1]Department of Fuel, Minerals & Metallurgical Engineering, Indian Institute of Technology (ISM) Dhanbad, India

*Corresponding Author Email: nitesh@iitism.ac.in

## Abstract

Numerical simulations were conducted to investigate the influence of inlet Reynolds number on the isothermal flow field in a lab-scale swirl combustor while keeping a fixed inlet swirl number of 0.67. The combustor geometry and baseline conditions were adopted from Taamallah et al. [1]. Unlike the experimental setup, which used axial-vane swirlers to generate rotation, this study imposed a velocity profile at the inlet to impart swirl. The simulations employed the Reynolds-averaged Navier–Stokes (RANS) approach with the shear stress transport (SST) $k-\omega$ turbulence model for closure, using ANSYS Fluent 2024R2 for all computations. A grid-independence study with meshes of approximately 0.4, 0.5, and 0.6 million elements showed that turbulent kinetic energy varied by less than 2% between the 0.5M and 0.6M grids, confirming adequate mesh resolution. The solver was validated against experimental data from Taamallah et al. [1], demonstrating close agreement in axial velocity distribution. The validated model was then used to simulate a higher inlet Reynolds number of ~30,000. Results showed contours of axial velocity and Centre-line profiles. An inner recirculation zone (IRZ), found by negative axial velocity in the core, formed in both cases and is crucial for flame stabilization. An outer recirculation zone (ORZ) was also seen near the expansion plane. An increase in the Reynolds number caused the peak forward axial velocity to rise by about 46.34%, while the reverse axial velocity at x = 0.10 m intensified by nearly 68%, showing stronger recirculation. Despite these changes in velocity magnitude, the axial location of the IRZ remained almost unchanged. If similar trends persist under reacting conditions, the flame's stabilization location (ignition kernel) would remain nearly unaffected by increases in Reynolds number, showing robust flame anchoring under varying inertial flows. Reacting flow simulations are planned for future work.

## 1. Introduction

Swirl-stabilized combustors are widely used in gas turbines and other propulsion devices due to their strong flame anchoring characteristics. The swirling motion induces central recirculation zones that help to maintain stable combustion by recirculating hot products and ignition sources upstream of the flame. Two recirculation regions are typically observed: an inner recirculation zone (IRZ) near the combustor axis and an outer recirculation zone (ORZ) near geometric expansions or walls. The IRZ is especially important for flame stabilization because the reversed flow provides hot burnt gases and radicals back toward the flame front [2,7]. Understanding how operating parameters such as the inlet Reynolds number affect these flow structures is therefore crucial for combustor design.

This work aims to quantify the sensitivity of the isothermal flow field in a lab-scale swirl combustor to changes in the inlet Reynolds number. The focus is on the aerodynamic aspects of the non-reacting isothermal flow, as a first step toward understanding flame behavior under different operating loads. The combustor geometry and baseline conditions are adopted from the experimental study of Taamallah *et al.*, but instead of physical vaned swirlers, we impose a pre-defined rotating velocity profile at the inlet. By comparing a baseline Reynolds number (~20,000) with a higher value (~30,000), we isolate the influence of inertial effects on recirculation zones and axial velocity patterns [3].

## 1.1. Literature Review

Swirling flows in combustion chambers have been the subject of extensive research due to their importance in enhancing flame stability. It is well established that a properly designed swirl introduces a strong low-velocity core flow (the IRZ) that acts as an ignition kernel by recycling hot products upstream of the flame. Taamallah *et al.* conducted an experimental and numerical study on a similar swirl combustor, focusing on the dynamics of helical vortex cores and their interaction with turbulent premixed flames. Their work highlighted how organized vortex structures contribute to flame holding and heat release distribution. Similar observations on the role of operating conditions in governing flow field transition were reported in tangentially fired gasifiers [4].

The present study builds on this context by examining the non-reacting counterpart: how the inlet Reynolds number modifies the purely aerodynamic flow features that ultimately underpin flame behavior.

Several computational and experimental investigations have explored isothermal swirl combustor flows. Common approaches involve specifying swirl through axial vanes or inlet guides and simulating the resulting flow using turbulence models. The shear-stress- transport (SST) $k - \omega$ model has often been employed for its reliability in capturing both near-wall behavior and free shear layers. While much of the existing literature emphasizes the flame response, there is consensus that the IRZ and ORZ shapes and magnitudes are sensitive to flow conditions [2,4]. However, detailed studies focused solely on the Reynolds number effect under constant swirl conditions are limited [3]. This gap motivates our investigation, which systematically isolates Reynolds number while holding the swirl generation mechanism fixed. The insights gained here can inform design margins for flame stability across different operating loads.

## 2. Methodology

### 2.1 Computational Modelling

The computational model is based on the laboratory-scale swirl combustor geometry reported by Taamallah et al. [1], as illustrated in Figure 1. The setup consists of a straight inlet pipe followed by a sudden expansion into a cylindrical chamber. Swirl is introduced at the inlet through a prescribed velocity boundary condition. The inlet section has a diameter of 0.038 m and a length of 0.045 m, while the main chamber measures 0.076 m in diameter and 0.225 m in length. The baseline operating conditions are summarized in Table 1, corresponding to an inlet Reynolds number of approximately 20,000.

The flow is modeled as steady, incompressible, and isothermal, using the Reynolds- averaged Navier–Stokes (RANS) framework with the shear-stress-transport (SST) $k - \omega$ turbulence model for closure. Simulations are carried out with the commercial solver ANSYS Fluent 2024R2.

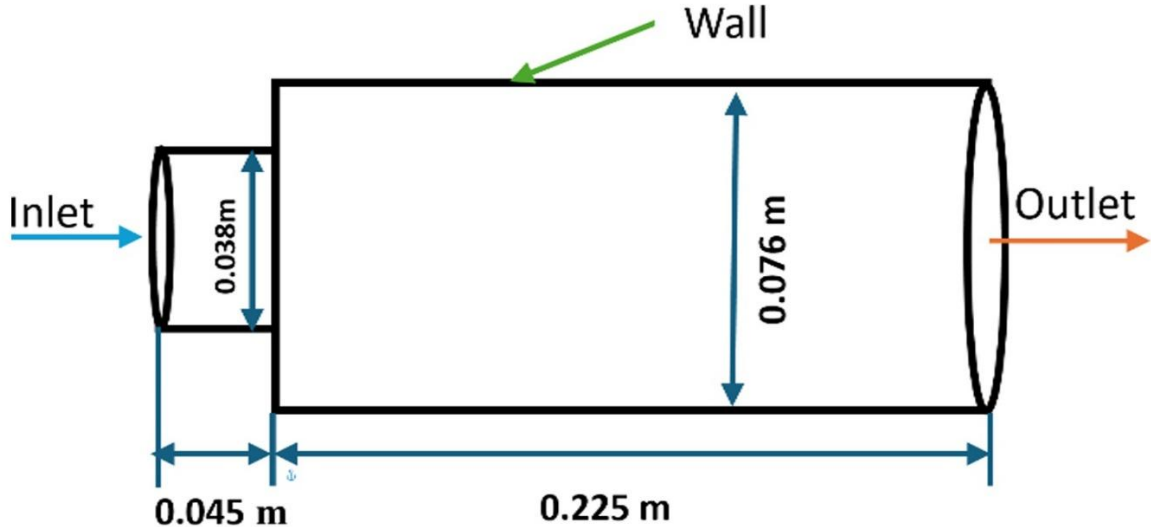

Fig. 1. Schematic of the combustor geometry (Taamallah *et al.* [1]).

**Table 1:** Operating conditions of the base case [1]

| Velocity (m/s) | Pressure (bar) | Temperature (K) |
|---|---|---|
| 8.2 | 1.0 | 300 |

## 2.2. Boundary Conditions

At the inlet plane, a velocity profile is imposed to provide the desired axial and tangential flow components. The mean inlet velocity is set such that the bulk Reynolds number is either approximately 20,000 of baseline case or 30,000 of higher case. The tangential velocity component is specified to replicate the same relative swirl number of 0.67 as in the baseline configuration of Taamallah *et al.*; The result is a swirling flow that enters the combustor, without explicitly modeling the swirler vanes. At the outlet plane, a static pressure boundary condition is applied typically zero-gauge pressure, allowing the flow to exit freely. All solid walls of the combustor are treated with no-slip boundary conditions. Since the focus is on isothermal flow, wall thermal conditions are not modeled. These boundary conditions ensure that the flow domain is closed properly and driven by the inlet momentum.

## 2.3. Solver Settings

The solver uses a coupled segregated algorithm. Pressure–velocity coupling is handled by the SIMPLE scheme. Gradients are evaluated by the least-squares cell-based method. Relaxation factors and residual monitors were used to ensure that the solution is fully converged. The SST $k-\omega$ model equations for turbulent kinetic energy ($k$) and specific dissipation rate ($\omega$) are solved along with the momentum equations. Because of the swirl, an elevated level of turbulence production is expected; the SST model helps limit excessive turbulent shear stress.

## 2.4. Grid Independence Study

A grid-independence study was performed with meshes of ~0.4, 0.5, and 0.6 million hexahedral elements for the base case at an inlet Reynolds number of ~20,000. The mesh is structured hexahedral in the central region and is grid-clustering to resolve the shear layers. A grid convergence study showed that most flow variables change minimally between 0.5M and 0.6M cells. In particular, the axial distribution of turbulent kinetic energy ($k$) along the centerline was monitored: Figure 2 shows that at 0.4 million cells the turbulent kinetic energy profile is slightly under-predicted, whereas the 0.5 million and 0.6 million grids produce virtually overlapping results. The turbulent kinetic energy effectively becomes grid-independent at ~0.5 million cells, showing adequate resolution of the shear-layer and recirculation structures. Therefore, the 0.5 million mesh was used for all subsequent simulations. The selected grid has an orthogonal quality of 0.52 and an aspect ratio of 6.7, as illustrated in Figure 3. The dimensionless wall distance ($y^+$) of the first grid elements is approximately 20, ensuring adequate near-wall resolution for the SST $k-\omega$ model turbulence model [5].

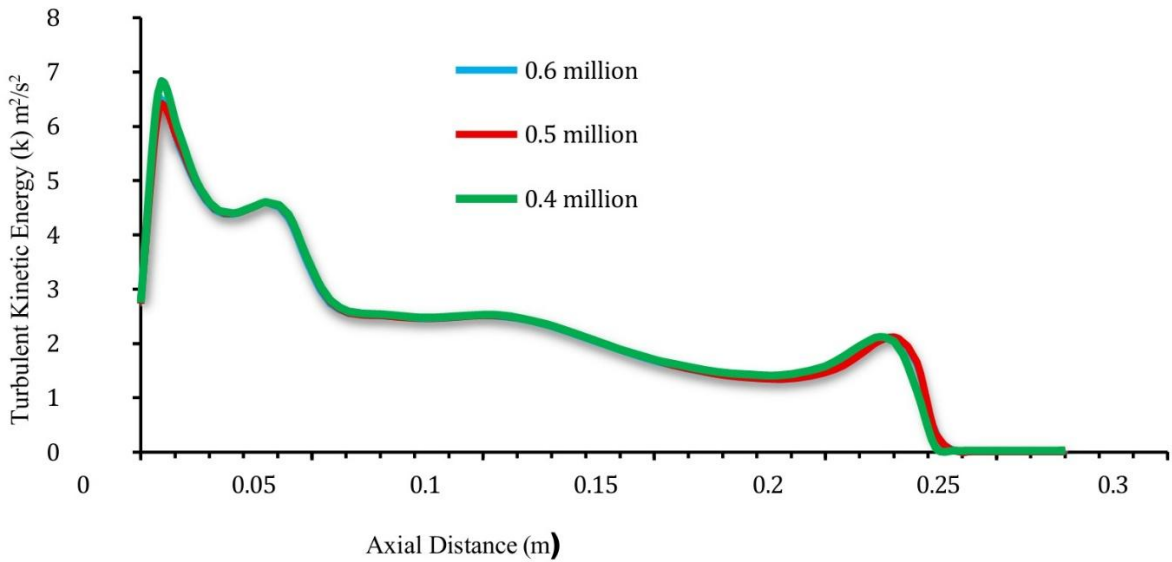

Fig. 2. Grid Independence Test

Additionally, the quantitative comparison of mesh sizes is summarized in Table 2. The variation in turbulent kinetic energy (TKE) between 0.5 M and 0.6 M cells was below 2%, confirming mesh independence. Hence, the 0.5 M cell mesh was adopted for all subsequent simulations.

Table 2. Grid-independence results in various mesh densities.

| Mesh size (Million cells) | Peak TKE (m²/s²) | % Change from finer grid | Conclusion |
|---|---|---|---|
| 0.4 | 6.81 | — | Under-predicted |
| 0.5 | 6.4 | -6.02% | Acceptable |
| 0.6 | 6.5 | +1.56% | Grid independent |

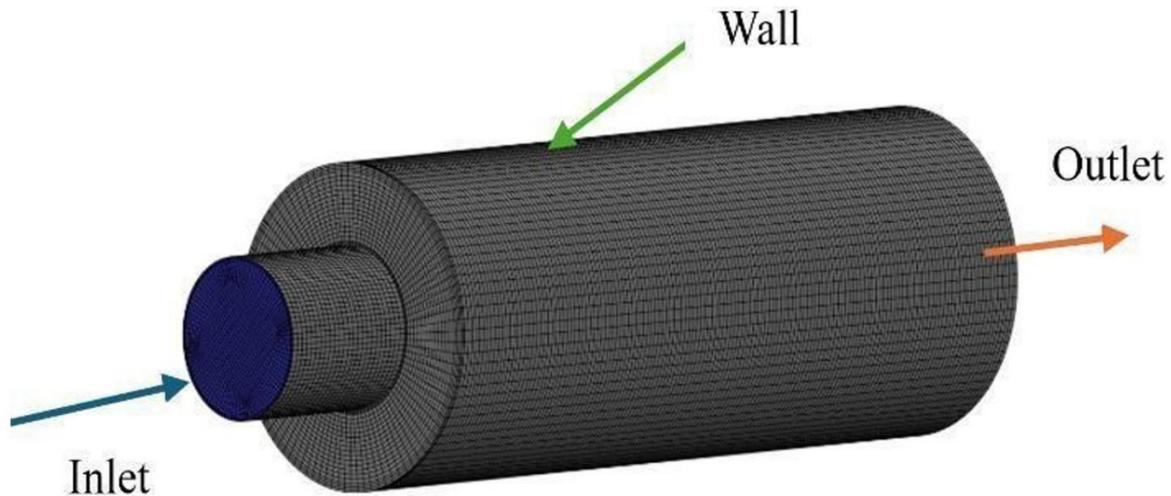

Fig. 3. Computational mesh with 0.5 million hexahedral elements.

## 2.5. Validation Study

The computational model was validated at Re ≈ 20,000 using the experimental dataset of Taamallah et al. [1], who reported time-averaged axial velocity profiles for the same geometry under non-reacting baseline conditions. For this purpose, the axial velocity distribution along the normalized radial direction at the downstream location x/R=0.5 was compared with the published measurements [1]. Similar cold-flow validation strategies for swirl burners have also been reported in the literature [6].

Figure 4 demonstrates that the numerical predictions are in close agreement with the experimental profile. Both the shape and magnitude of the simulated velocity field align well with the reported data, indicating that the SST turbulence model successfully reproduces the core flow and shear-layer behavior.

Small deviations are observed, which can be attributed to model simplifications such as neglecting swirl vanes and combustion effects. Overall, the solver validation against experimental data shows that the RANS–SST framework can capture the key flow features in the studied combustor with satisfactory accuracy.

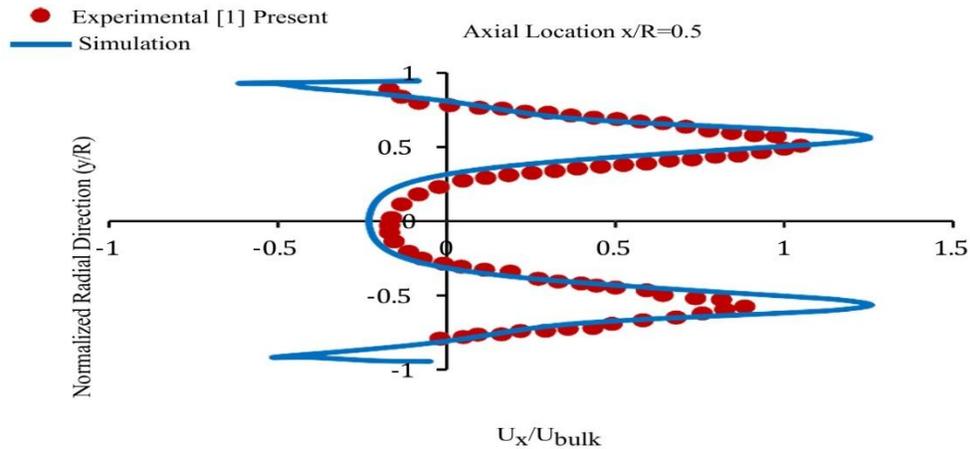

Fig. 4. Comparison of radial profiles of axial- velocity with experimental data [1] at x/R = 0.5.

## 3. Results and Discussion

### 3.1. Axial Velocity Contours

Figure 5 shows the axial velocity fields for Re ≈ 20,000 and Re ≈ 30,000. Both cases show a central inner recirculation zone (IRZ) along the axis and a weaker outer recirculation zone (ORZ) near the wall. Increasing Reynold number strengthens the core jet but does not significantly alter the position or intensity of the recirculating zones.

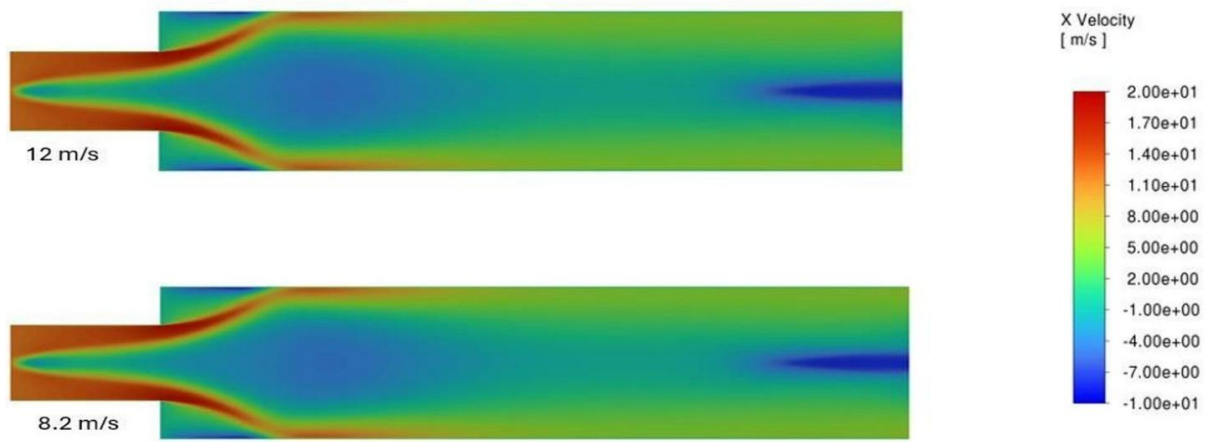

Fig. 5. Contours of axial velocity for Re = 30000 top, Re = 20000 bottom

### 3.2. Axial Velocity Profiles

Centerline velocity distributions (Figure 6) confirm the contour observations. In both cases, the IRZ strength remains similar, with its extent only shifting slightly downstream at higher Reynolds number. The IRZ extends slightly further downstream at higher Reynolds number, but its magnitude is insensitive to inlet conditions. Outside the IRZ, the higher-Re case produces a steeper velocity recovery and larger downstream peak velocity, reflecting the stronger jet. The overall profile shapes remain consistent, with only minor variations due to the increased flow rate.

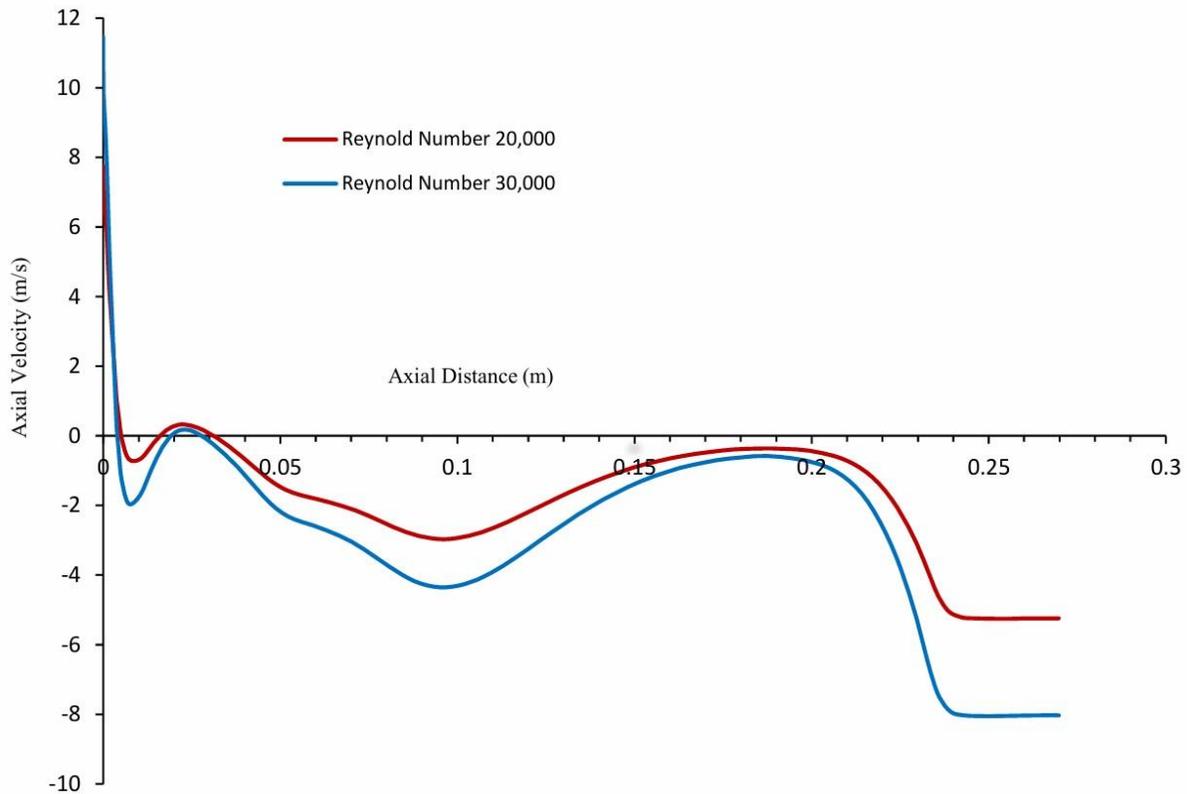

Fig. 6. Axial Variations of axial velocity for the cases investigated with two different Reynolds numbers.

### 3.3. Quantitative Comparison of Reynolds Number Effects

To enable a clear comparison between the two Reynolds number conditions, the key flow parameters were extracted and are summarized in Table 3. The peak forward axial velocity increased by approximately 46.34% when the Reynolds number was raised from 20,000 to 30,000, demonstrating a significantly stronger central jet. Despite this increase in forward momentum, the inner recirculation characteristics remained stable. The axial reverse velocity at x = 0.10 m became 68% more negative, indicating enhanced recirculation strength at higher Reynolds number. Overall, the IRZ location remained nearly unchanged, confirming that the fundamental recirculation structure is largely insensitive to inlet inertia, even though vortex intensity increases with Reynolds number.

**Table 3.** Comparison of major flow parameters at two Reynolds numbers

| Parameter | Re = 20,000 | Re = 30,000 | Difference | Observation |
|---|---|---|---|---|
| Peak axial velocity (m/s) | +8.2 | +12 | +46.34 % | Stronger jet |
| Axial velocity at x = 0.10 m (m/s) | –2.5 | –4.2 | 68% | Stronger reverse flow at higher Re |
| Recirculation strength | Moderate | Strong | Stronger | Higher Re increases vortex strength. |

### 3.4 Streamline and Vorticity Analysis

Figure 7 illustrates the streamline patterns for both Reynolds numbers. The results show the formation of a well-defined inner recirculation zone (IRZ) and a weaker outer recirculation zone (ORZ) near the expansion region. The vortex core remains nearly unchanged in position, confirming strong flow stability despite increased inlet momentum.

Figure 8. represent vorticity contours at inlet velocities of 8.2 m/s and 12 m/s. Higher inlet velocity produces stronger shear-layer vorticity near the jet exit and a more pronounced downstream negative-vorticity region, indicating a longer recirculation bubble. The intensified vorticity at 12 m/s also reflects increased jet penetration, enhanced mixing, and the appearance of slight flow asymmetry due to higher instability.

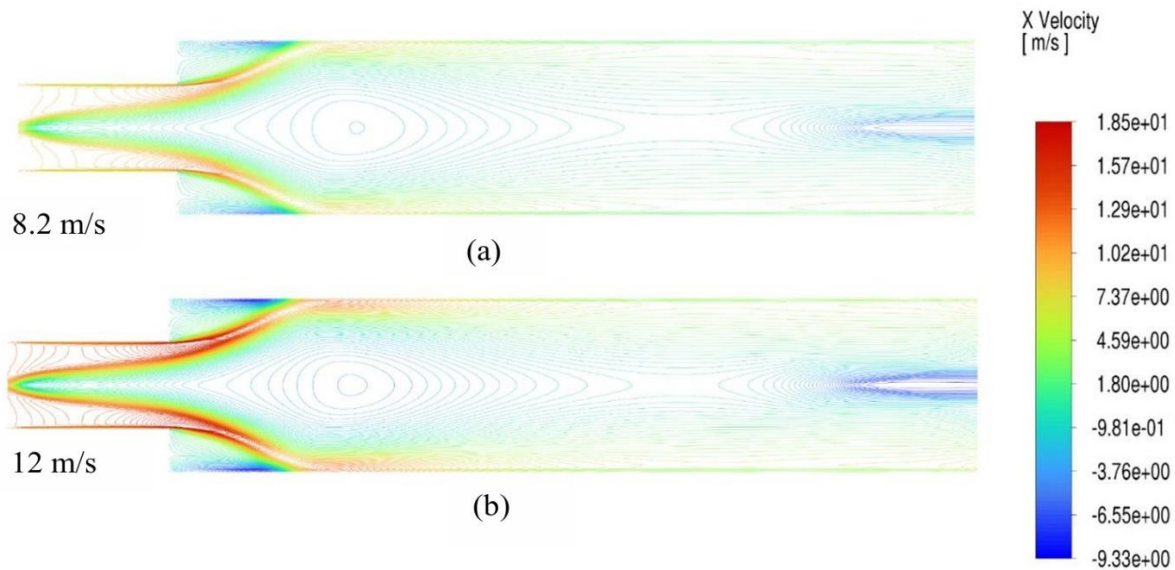

**Fig. 7.** Streamline contours showing recirculation patterns at (a) Re = 20,000 and (b) Re = 30,000.

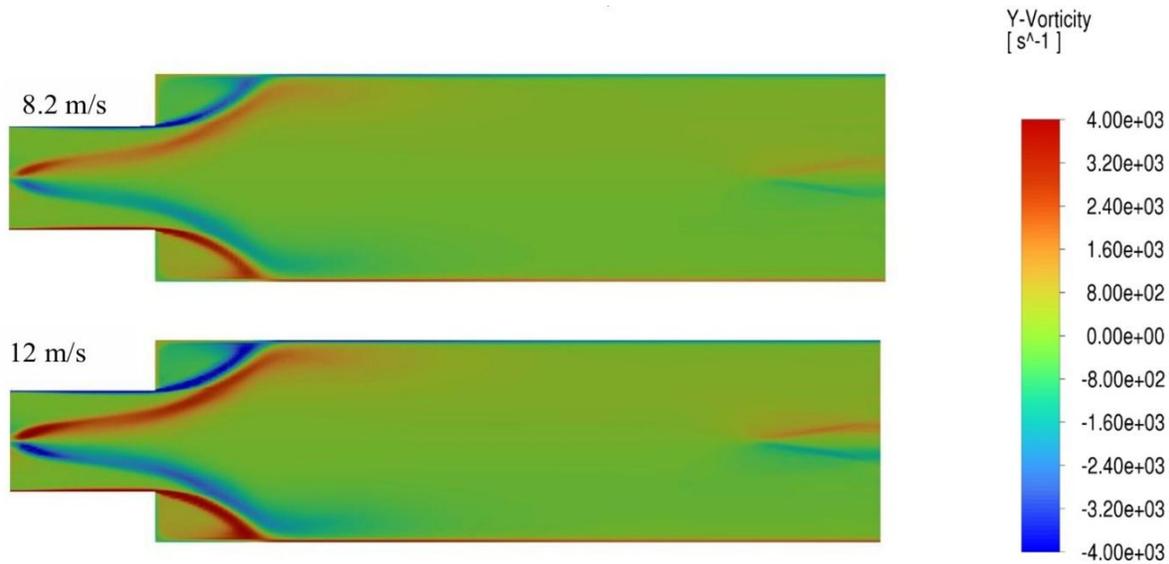

**Fig. 8.** Vorticity magnitude distribution for both Reynolds numbers 20000 and 30000.

### 3.5. Discussion of Reynolds Number Effects

The results show that the IRZ is a stable and robust feature, largely unaffected by changes in Reynold number, consistent with prior observations of swirl-stabilized flows [6,7]. This implies that flame stabilization through central recirculation would persist under varying operating loads. While higher inlet momentum enhances core velocity and may influence mixing and pressure loss, these aerodynamic changes do not compromise the availability of recirculation for flame holding. The present non-reacting study therefore sets up a reliable aerodynamic baseline, with future reacting-flow simulations needed to assess heat- release effects [2,8]

### 4. Conclusion

A numerical study was performed to assess how the inlet Reynolds number affects the flow in a lab-scale swirl-stabilized combustor. Using a RANS–SST $k - \omega$ model in Fluent and an inlet velocity profile to generate swirl, two cases were compared: a baseline Re ≈ 20,000 and a higher Re ≈ 30,000. A grid-independence study confirmed that a mesh of about 0.5 million cells suffices to capture the flow features. The solver predictions were validated against experimental data, showing good agreement in the size of the recirculation zones and velocity magnitudes. In the results, both cases exhibited a strong inner recirculation zone (IRZ) near the axis and an outer recirculation zone (ORZ) near the expansion plane. When the Reynolds number was increased, the peak forward axial velocity rose by about 46.34%, and the reverse velocity within the IRZ strengthened by nearly 68%. However, the axial position of the IRZ remained almost unchanged.

These findings imply that the flame-anchoring recirculation zone is insensitive to changes in inlet inertia. Accordingly, one would expect the flame stabilization point (ignition kernel) to stay effectively fixed as the flow rate varies. In other words, the combustor should maintain robust flame anchoring under different inlet Reynolds numbers. Reacting flow simulations to test this hypothesis are planned for future work.

## Acknowledgements

The present work is supported by the Faculty Research Scheme (FRS) Grant; FRS (217)/2024-25/FMME of IIT-ISM Dhanbad.

**Author Biographies**

**Madan Lal Mahato**

Madan Lal Mahato is a PhD scholar in the Department of Fuel, Minerals & Metallurgical Engineering at the Indian Institute of Technology (ISM) Dhanbad, India. His research focuses on advanced combustion systems, numerical modeling of swirl-stabilized combustors, and energy conversion technologies. He has experience with computational fluid dynamics (CFD) using ANSYS Fluent for turbulent reacting and non-reacting flows. His current work investigates flame stabilization mechanisms under varying operating conditions with emphasis on clean and efficient energy systems.

**Nitesh Kumar Sahu**

Dr. Nitesh Kumar Sahu is an Assistant Professor in the Department of Fuel, Minerals & Metallurgical Engineering at the Indian Institute of Technology (ISM) Dhanbad, India. His research interests include combustion, gasification, entrained bed systems, and computational modeling of multiphase flows. He has published in reputable international journals and conferences, contributing to the understanding of combustion and clean energy technologies. Dr. Sahu guides several post graduate students in energy and combustion research.

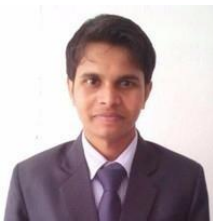

Madan Lal Mahato
PhD scholar
Fuel, Minerals & Metallurgical Engineering
Indian Institute of Technology (ISM)
Dhanbad

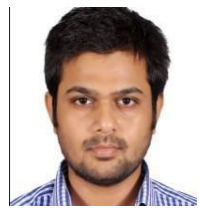

Dr. Nitesh Kumar Sahu
Assistant Professor
Fuel, Minerals & Metallurgical Engineering
Indian Institute of Technology (ISM) Dhanbad.

International Conference on Innovations in Fluid Power: Shaping the Future of Mobile and Industrial Machinery
Department of Mechanical Engineering, IIT (ISM) Dhanbad, India
23–24 January 2026